\pdfoutput=1

\documentclass[11pt]{article}
\usepackage{jheppub}

\usepackage{amsmath}
\usepackage{latexsym,amssymb}
\usepackage{graphicx,color,slashed}
\usepackage{bbm} 
\usepackage{mathtools}
\usepackage{ifpdf}

\newcommand{\cA}{\mathcal{A}}

\newcommand{\cF}{\mathcal{F}}
\newcommand{\cG}{\mathcal{G}}
\newcommand{\cH}{\mathcal{H}}
\newcommand{\cL}{\mathcal{L}}
\newcommand{\cM}{\mathcal{M}}
\newcommand{\cN}{\mathcal{N}}
\newcommand{\cO}{\mathcal{O}}

\newcommand{\cP}{\mathcal{P}}
\newcommand{\cT}{\mathcal{T}}

\newcommand{\mb}{\mathbf}

\newcommand{\be}{\begin{equation}}
\newcommand{\ee}{\end{equation}}
\newcommand{\ba}{\begin{eqnarray}}
\newcommand{\ea}{\end{eqnarray}}
\newcommand{\nn}{\nonumber}

\newcommand{\N}{\mathcal{N}}

\def\E{{$E_{7(7)}$}}

\newcommand{\rf}[1]{(\ref{#1})}
\newcommand{\bea}{\begin{eqnarray}}
\newcommand{\eea}{\end{eqnarray}}

\def\bfzero{\relax{\rm I\kern-.18em 0}}
\def\bfone{\relax{\rm 1\kern-.35em 1}}
\def\twomat#1#2#3#4{\left(\begin{array}{cc}
\end{array}
\right)}

\newcommand{\mc}[1]{\mathcal{#1}}

\newcommand{\f}{{\bf f}}
\newcommand{\df}{\bar{\bf f}}
\newcommand{\D}{\Delta}
\newcommand{\M}{{\bf M}}

\newcommand{\ph}{\phantom}

\title{On quantum compatibility of counterterm deformations and duality symmetries in
$\cN\geq 5$ supergravities}

\author[a]{Renata Kallosh,}
\author[b]{Hermann Nicolai,}
\author[c]{Radu Roiban,}
\author[a]{and Yusuke Yamada}
\affiliation[a]{Stanford Institute for Theoretical Physics and Department of Physics, Stanford University, Stanford, CA 94305, USA}
\affiliation[b]{Max-Planck-Institut f\"ur Gravitationsphysik (Albert-Einstein-Institut) M\"uhlenberg 1, D-14476 Potsdam, Germany}
\affiliation[c]{Institute for Gravitation and the Cosmos,
The Pennsylvania State University, University Park, PA 16802, USA}
\emailAdd{kallosh@stanford.edu}
\emailAdd{nicolai@aei.mpg.de}
\emailAdd{radu@phys.psu.edu}
\emailAdd{yusukeyy@stanford.edu}

\notoc

\abstract{In ${\cal N}=5, 6, 8$ supergravities there are hidden  symmetries of equations of motion, described by  duality groups $SU(1,5), \, SO^*(12), \, E_{7(7)}$ respectively.  UV divergences and known candidate counterterms violate the deformed duality symmetry current conservation. Extra higher derivative terms in the action are required to restore duality.   We study the effect of a two-vector part of the counterterm for  $\cN\geq 5$ supergravities using the universality of the symplectic structure of extended supergravities.  We construct a compact form of  a deformed action with infinite number of higher derivative terms and restored duality symmetry with deformation parameter $\lambda$.  We find, in $\lambda^2$ approximation, that the $SU(\cN)$ symmetry of the deformed theory is restored on shell.
}

\begin{document}

\maketitle

\newpage

 \tableofcontents{}

\newpage

%%%%%%%%%%%%%%%%%%%%%%%%%%%%%%%%%%%%%%%%%%%%%%%%%%%%%%%%%%%%%%%%%%%%%%
%%%%%%%%%%%%%%%%%%%%%%%%%%%%%%%%%%%%%%%%%%%%%%%%%%%%%%%%%%%%%%%%%%%%%%
%%%%%%%%%%%%%%%%%%%%%%%%%%%%%%%%%%%%%%%%%%%%%%%%%%%%%%%%%%%%%%%%%%%%%%
%%%%%%%%%%%%%%%%%%%%%%%%%%%%%%%%%%%%%%%%%%%%%%%%%%%%%%%%%%%%%%%%%%%%%%
%%%%%%%%%%%%%%%%%%%%%%%%%%%%%%%%%%%%%%%%%%%%%%%%%%%%%%%%%%%%%%%%%%%%%%

\section{Introduction}
All classical extended supergravities with $\cN$ local supersymmetries have  duality symmetry, as shown by Gaillard and Zumino
\cite{Gaillard:1981rj}. These symmetries rotate equations of motion into Bianchi identities and are consistent with local extended supersymmetry of the classical action. 

The local UV divergences at the loop level can be eliminated (absorbed into a redefinition of parameters)  if the classical action of extended supergravities are deformed, to preserve a duality symmetry in presence of higher derivative terms. The issue of compatibility of the deformed duality symmetric extended supergravity with a global $\cN$-extended supersymmetry of the on-shell amplitudes will be addressed here.

A deformation of $\cN = 8$ supergravity by the candidate counterterms (CTs) \cite{Kallosh:1980fi}, \cite{Howe:1980th} leads to a 
violation of the duality current conservation \cite{Kallosh:2011dp}, \cite{Kallosh:2011qt},  unless the consistent procedure of the deformation of the twisted selfduality condition \cite{Bossard:2011ij},\cite{Carrasco:2011jv} can be implemented. 
%A  general procedure  in a form proposed in \cite{Carrasco:2011jv} 
In its general form proposed in \cite{Carrasco:2011jv}  it has been already applied for  Born-Infeld models with higher derivatives with $U(1)$ duality in \cite{Chemissany:2011yv}. 
Other examples of the restoration of duality current conservation with rigid $\cN=2$ supersymmetry and  $U(1)$ duality  were presented in \cite{Broedel:2012gf}. The procedure of \cite{Carrasco:2011jv} was however not explicitly applied to extended supergravities.

We will  solve the first part of the problem here, in a particular sector of the theory: we will construct a deformed bosonic action of $\cN$-extended supergravity where a two-vector part of the CT is added to the classical action.  All higher order terms with higher and higher derivatives will be identified, so that the deformed actions in ${\cal N}=5, 6, 8$ supergravities in the two-vector sector have  restored $SU(1,5), \, SO^*(12), \, E_{7(7)}$ duality symmetry, respectively. We will investigate the properties of the deformed bosonic action here and  study the supersymmetric embedding of the deformed action and the superamplitudes.

A  consistent reduction of $\cN=8$ to  all pure extended supergravities allows us  to work with all  $\cN\geq 4$ models.   The $\cN= 4$ pure supergravity has a $U(1)$ duality anomaly \cite{Marcus:1985yy}, \cite{Carrasco:2013ypa}, which might have caused the four-loop UV divergence \cite{Bern:2013uka}. It was suggested recently in \cite{Bern:2017rjw} 
that the one-loop anomalous amplitudes in this theory can be cancelled by a finite local counterterm. It remains an interesting open problem to understand the consequences of this
(and perhaps higher-loop) counterterm(s) on the four-loop divergence of the four-graviton amplitude.

Meanwhile, the four-loop UV divergence  in $\cN= 5$ supergravity is absent, \cite{Bern:2014sna}. Moreover, it has been recently established in \cite{Freedman:2017zgq} that $\cN\geq 5$ supergravities do not exhibit $U(1)$ duality anomalies in their one-loop amplitudes, of the kind known to be present in $\cN= 4$ case \cite{Carrasco:2013ypa}.

The first relevant prediction of a UV finiteness of $\cN=8$  supergravity due to  \E\,  symmetry in \cite{Kallosh:2011dp}  was based on an observation that 
 the Lorentz and SU(8) covariant, \E  \, invariant unitarity constraint expressing the 56-dimensional \E \, doublet via 28 independent vectors, and consistent with supersymmetry,  is unique \footnote{In case of $\cN=8$  supergravity the direct and simple finiteness argument is based on the absence of light-cone supersymmetric invariant counterterm candidates \cite{Kallosh:2009db}. We are grateful to L. Brink for a recent reminder that light-cone CT's are still not available, despite a significant  effort. But since here we are interested also in 
 $  5 \leq \cN <8$ supergravity we cannot rely on  light-cone superspace, which is known only for $\cN=8$  supergravity. We will work in the Lorentz covariant approach and try to use the duality/supersymmetry argument.}. This argument is easy to    extended  to other cases of $\cN\geq 5$ supergravities since it is based on the geometric nature of the ${\cG\over \cH}$ coset space where scalars are coordinates. Later on  in \cite{Kallosh:2012yy}  the argument was given that for all $\cN\geq 5$ supergravities the procedure of restoration of the duality current conservation broken by the CT  is not available. The argument in \cite{Kallosh:2012yy} was based on the properties of invariants  of the groups of type $E_7$, which are duality groups in $\cN\geq 5$ supergravities \cite{Ferrara:2011dz}. It suggested that a deformation of the twisted selfduality condition for groups of the type $E_7$, consistent with supersymmetry,  is not possible. Additional reasons for  an obstruction to \E deformations in  $\cN=8$  supergravity based on superconformal $SU(2,2|8)$  algebra were developed in \cite{Gunaydin:2013pma}.

 It is important to stress here that the conjectured breaking of continuous \E\ to discrete \E($\mathbb{Z}$)  would be a non-perturbative effect, whereas we are 
 analyzing here only perturbative supergravity. 
The perturbative quantization of $\cN = 8$ supergravity is studied in  \cite{Bossard:2010dq} in a formulation where its $E_{7(7)}$ symmetry is realized off-shell, 
but Lorentz invariance is no longer manifest. 

Relying on the cancellation of $SU(8)$ current anomalies it is shown there that there are no anomalies for the non-linearly realized \E\, either. As a consequence, 
the \E\,  Ward identities can be consistently implemented and imposed at all orders in perturbation theory, and therefore potential divergent CTs must respect 
the full non-linear  \E\, symmetry. 

In view of the highly non-trivial cancellation of the UV divergences in $\cN=5$ supergravity in four loops discovered in \cite{Bern:2014sna} and the fact that no new explanations of this fact, besides the one in  \cite{Kallosh:2012yy}, have been suggested we would like to revisit and clarify the status of the duality conservation arguments in \cite{Kallosh:2011dp}, \cite{Kallosh:2011qt} and   \cite{Bossard:2011ij}.

The UV finiteness of $\cN=5$ supergravity in four loops established in  \cite{Bern:2014sna} may shed some light on the UV properties of the maximal $\N=8$ supergravity, if there exists a universal formalism describing all $\cN$-extended supergravities. Such a formalism is, indeed, available for $\cN\geq 2$ and it was constructed to describe the supersymmetric black hole universality \cite{Ferrara:1996um}, \cite{Ferrara:2006em}. In $\cN= 2$ the  special geometry is represented by a symplectic section \cite{deWit:1984wbb}, \cite{Strominger:1990pd}. The symplectic sections for higher $\cN$ have been constructed  in \cite{Andrianopoli:1996ve}, \cite{Andrianopoli:2006ub}.

We would also like to briefly comment on the very recent 5-loop
calculation~\cite{Bern:2018jmv} demonstrating the presence of a divergence in $N=8$
supergravity for $D_{crit} = \frac{24}5$, and thus the absence of
%`miraculous cancellations' 
enhanced cancellations \cite{Bern:2017lpv}
at least in that case. Although this result
may be interpreted as a hint that in four dimensions $\cN=8$ supergravity
might diverge at seven loops it should be emphasized that the question of
finiteness (or not) of $\cN=8$ supergravity remains wide open. The approach
taken in this paper relies essentially on the exceptional symmetry $E_7$
which has no analog in fractional critical dimensions, and on the fact
that any CT must respect an extension of the exceptional duality
symmetry, along the lines of the construction done in this paper. The
question of whether a higher order CT exists that is both fully
supersymmetric and fully duality invariant thus remains a challenge on
a par with an explicit calculation at seven loops.

Our purpose here is to make an analysis using the relatively simple two-vector sector of the theory, extending 
earlier results of ref.~\cite{Bossard:2011ij}. After the deformed bosonic action with duality symmetry will be 
presented we will study its supersymmetric embedding. 

\section{ Twisted selfduality constraint and its deformation in $\N\geq 5$ supergravities}
\label{TDG}

The models of ${\cal N}=5, 6, 8$ supergravities are reviewed in detail in Appendix~\ref{gen4d}, based on \cite{Andrianopoli:1996ve}, \cite{Andrianopoli:2006ub}.   The scalars are coordinates of the ${\cG\over \cH}$ cosets, see Table 1;   the notation is universal for all of them. Their duality groups $\cG$ are $SU(1,5), \, SO^*(12), \, E_{7(7)}$ respectively.  The isotropy groups $\cH$ are $U(5), U(6), SU(8)$, respectively.

We are looking at the bosonic part of the two-vector sector of the CT  \cite{Kallosh:1980fi}, \cite{Howe:1980th}, which has a manifest duality symmetry as well as a supersymmetry, under condition that all fields in the CT satisfy classical equations of motion, ${\delta S_{\rm cl}\over \delta \phi}=0$. But once such a CT is added to the action with some constant $\lambda$ in front of it, the new equation of motion has a correction
\be
{\delta S_{\rm deformed}\over \delta \phi} = {\delta S_{\rm cl}\over \delta \phi} + \lambda {\delta S_{CT}\over \delta \phi} =0.
\ee
In particular, the $\lambda$-dependent terms  break  duality current conservation \cite{Kallosh:2011dp}, \cite{Kallosh:2011qt} at order ${\cal O}(\lambda^2)$. 
New terms of ${\cal O}(\lambda^2)$ are therefore necessary to correct this issue; they, in turn, push the non-conservation of duality current to ${\cal O}(\lambda^3)$, etc. 
The current conservation is  restored with an infinite number of higher order terms, which also have an infinite number of higher derivatives, \cite{Bossard:2011ij},\cite{Carrasco:2011jv}.

We will now study dualities in ${\cal N} \ge 5$ supergravities, \cite{Andrianopoli:1996ve}, \cite{Andrianopoli:2006ub}; the field content of these theories is given only by the corresponding 
gravitational multiplet. In the case of $\cN=4$ supergravity the duality symmetry is anomalous, \cite{Marcus:1985yy}, \cite{Carrasco:2013ypa} but $\N\geq 5$ are anomaly-free \cite{Freedman:2017zgq}.
These theories contain in the bosonic sector
 the metric, a number $n_v$ of vectors and $m$ of (real)
scalar fields, see Table \ref{topotable}. The relevant classical  vector and scalar part of action has the
following general form:
\begin{equation}
{\cal L}_{vec} \,  = \, {\mbox{i}} \, \left [  {\bar {\cal
N}}_{\Lambda \Sigma}  F_{\mu\nu}^{-\Lambda } F^{- \Sigma |\mu\nu} -   {\cal N}_{\Lambda \Sigma}  F_{\mu\nu}^{+\Lambda } F^{+ \Sigma |\mu\nu} \right] +\frac 12 \,g_{rs}(\Phi) \partial_{\mu}
\Phi^{r}\partial^{\mu}\Phi^{s}\,, \label{lagrapm}
\end{equation}
where $g_{rs}(\Phi)$ ($r,s,\cdots =1,\cdots ,m$) is the scalar metric on the scalar manifold ${\cal M}_{scalar}$ of real dimension
 $m$ and the vectors kinetic matrix
${\cal N}_{\Lambda\Sigma}(\Phi) $ is a complex, symmetric, ${n_v} \,
\times \, {n_v}$ matrix depending on the scalar fields, see Table \ref{topotable}. $F^{\pm \Lambda}$ are self-dual and anti-self-dual combinations of the vectors field strengths (see  
Appendix~\ref{gen4d} for details).
{\footnotesize
\begin{table}[h]
\begin{center}
\caption{\sl Scalar Manifolds of $\cN\geq 4$ Extended Supergravities}
\label{topotable}
\begin{tabular}{|c||c|c|c|c|c| }
\hline N &  Duality group $\cG$ & isotropy $\cH$ & ${\cal M}_{scalar}$ & $n_v $&$ m$   \\
\hline \hline
\hline $4$  &   $SU(1,1)\otimes SO(6)$ & $U(4) $ &
$\frac{SU(1,1)}{U(1)}$  & $6$& $2$ \\
\hline $5$  &  $SU(1,5)$ &$U(5)$  & $\frac{SU(1,5)}
{S(U(1)\times U(5))}$ & 10& 10 \\
\hline $6$  &  $SO^\star(12)$ &$U(6)$ &
$\frac{SO^\star(12)}{U(1)\times SU(6)}$ & 16& 30 \\
\hline $7,8$&  $E_{7(7)}$ & $SU(8)$ &
$\frac{ E_{7(7)} }{SU(8)}$ & 28& 70 \\
\hline
\end{tabular}
\end{center}
{ In the table, $n_v$ is the number of vectors and $m$ is the number of real scalar fields. In all the cases the duality
group $\cG$ is embedded in ${Sp}(2\,n_v,\mathbb{R})$.}
\end{table}
}

The formalism of symplectic sections \cite{Andrianopoli:1996ve}, \cite{Andrianopoli:2006ub} corresponds to  a particular parametrization of the coset representative. It
allows a better way to study duality symmetry of extended supergravities for the case of a general $\cN$. The details are in Appendix~\ref{gen4d} for $\cN=5,8$\footnote{In case of $\cN=6$ the details of the coset space ${SO*(12)\over U(6)}$ are given in \cite{Andrianopoli:1996ve,Andrianopoli:2006ub}. There are 16 graviphotons, 15 in the twice-antisymmetric representation of U(6) plus a singlet, see~\cite{Andrianopoli:1996ve}, p.17 or  \cite{Andrianopoli:2006ub} p. 71-72. These subtleties do not affect our analysis.} and  we give examples of symplectic sections in $\cN=5$ and $\cN=8$ supergravity in Appendix \ref{EGsymplecticsections}. Instead of a metric
${\cal N}_{\Lambda\Sigma}(\Phi) $ in the vector space,  in eq.~\rf{lagrapm} one can introduce duality doublets -- referred to as a symplectic section --  depending on scalars of the theory 
 \begin{equation}
\left(\begin{array}{c} f^\Lambda{}_{AB} \cr h_{\Lambda}{}_{AB}\cr \end{array}\label{ss}\right)\end{equation}
so that  the kinetic matrix $\cN$ can be written in terms of the sub-blocks ${\bf f}$, ${\bf h}$
as $\cN= {\bf h}\,{\bf f}^{-1}$ or component-by-component as
\begin{equation}
\cN_{\Lambda \Sigma} = { h}_{\Lambda AB}\,({ f}^{-1})^{AB}{}_ {\Sigma} \ .
\label{nfh}
\end{equation}
The vector doublet is defined by the vector field strength $F^\Lambda_{\mu \nu}  \equiv  {\frac{1}{2 }} \,\left (
\partial_{\mu } A^\Lambda_{\nu} \, - \, \partial_{\nu }
A^\Lambda_{\mu} \right )$ and by the derivative of the action over it,  namely, $ {{}^\star G}_{\Lambda|\mu\nu} \, \equiv \,  {\frac 12 } { \frac{\partial
{\cal L}}{\partial  F^\Lambda_{\mu\nu}}}$
\begin{equation}
{ \mathcal{F}}  \, \equiv \, \left(
  \begin{array}{c}
  { F}^{\Lambda} \cr  { G}_{\Lambda} \cr  
  \end{array}
\right)\, .
\label{sympvec1}\end{equation} 
{ The only way to construct  $\cG$-invariants is by contracting the symplectic doublets}. For example, 
the graviphoton -- the $\cN(\cN-1)/2$-component supersymmetric partner of the $\cN$-component gravitino $\psi_A$ -- is defined as
\be
T_{AB}^{\pm}=(f^\Lambda{}_{AB}, h_{\Lambda AB}) \left(\begin{array}{cc} 0& -\bfone \cr \bfone & 0\cr
\end{array} \right)\, \left(\begin{array}{c} F^{\pm \Lambda} \cr G^{\pm}_\Lambda\cr \end{array}\right).
\label{inv}\ee
Here $ F^{\mp \Lambda}_{\mu\nu}$ are the Maxwell field strength in the action in eq. \rf{lagrapm} , whereas $G^{\mp }_{\Lambda |\mu\nu}$
are defined as derivatives of the action over $ F^{\mp \Lambda}_{\mu\nu}$,
\begin{equation}
G^{\mp }_{\Lambda |\mu\nu} \, \equiv \, \mp{\frac {\rm i}{2} } {
\frac{\partial {\cal L}}{\partial  F^{\mp \Lambda}_{\mu\nu}}}\,.
\label{gtensor}
\end{equation}
Note that the graviphoton is a $\cG$-invariant and are covariant under the $\cH$-symmetry, the $U(\cN)$ for $6\geq\cN\geq 4$ and $SU(8)$ for $\cN=8$.

We consider a  two-vector part of the CT in
 $\N\geq 5$ supergravities, \cite{Kallosh:1980fi}, \cite{Howe:1980th}. The relevant expression, a supersymmetric partner of $D^{2k}R^4$,  
 depends on the graviphoton $T_{\mu\nu AB}^-$  and its conjugate defined in eqs. \rf{inv}, \rf{gravi}:
  \be
{\cal L}_{CT}= \lambda \,   T^{- }_{ AB} \,  \Delta \, \bar  T^{- \,  AB} \ .
\label{ct}
\ee
The simplest case of the $R^4$ CT is
\be 
{\cal L}_{CT}=\lambda \cT ^{\alpha\beta\gamma\delta\dot{\alpha}\dot{\beta}
\dot{\gamma}\dot{\delta}}  \nabla_{\alpha\dot{\delta}} 
 T_{\beta\gamma AB} \nabla_{\delta\dot{\alpha}} \bar {T}_{\dot{\beta}\dot{\gamma}}^{ AB} 
\label{la}\ee
with $\cT ^{\alpha\beta\gamma\delta\dot{\alpha}\dot{\beta}
\dot{\gamma}\dot{\delta}}  = \lambda \, C^{\alpha\beta\gamma\delta} \bar C^{\dot{\alpha}\dot{\beta}
\dot{\gamma}\dot{\delta}}  $
%, $C^{\alpha\beta\gamma\delta}$ 
being the Bel-Robinson tensor in spinor notation and
$ \nabla_{\alpha\dot{\delta}} $ an $\cH$-covariant space-time derivative.  
In the $R^4$ case the explicit differential operator in eq. \rf{ct} acting on two-forms $f_{\mu\nu}$ is defined as follows \be\label{DiffOp}
(\Delta (f))_{\mu\nu} \equiv
\Delta_{\mu\nu}{}^{\rho\sigma} f_{\rho\sigma}
:= \nabla_\kappa \cT_{[\mu}{}^{\kappa\lambda[\sigma}  
\nabla_\lambda \delta_{\nu]}^{\rho]} f_{\rho\sigma}\; ;
\ee 
it maps a self-dual 2-form into an anti-self-dual one and vice versa.
Here  the Bel--Robinson tensor is given in the vector form
\be 
\cT^{\mu\nu\sigma\rho} \equiv C^{\mu\kappa\sigma\lambda} C^\nu{}_\kappa{}^\rho{}_\lambda 
- \frac{3}{2} g^{\mu[\nu} C^{\kappa\lambda]\sigma\vartheta} C_{\kappa\lambda}{}^\rho{}_\vartheta \, , 
\ee
with the Weyl tensor $C_{\mu\nu\sigma\rho}$. 
The $\cH$-covariant  field strength of the graviphoton is 
$T_{  \alpha\beta\, AB} \equiv \sigma^{\mu\nu}_{\alpha\beta} T^{AB}_{\mu\nu}$ 
and its complex conjugate is 
$\bar{T}_{\dot\alpha\dot\beta AB}$. 

Note that as it is known from \cite{Brodel:2009hu, Elvang:2010kc}, the $R^4$ CT does not have a supersymmetric completion that is also invariant under a duality group with real coefficients. It is however believed that its nonperturbative completion is invariant under a duality group with integer coefficients \cite{Green:2010wi}.

For  an $D^{2k}R^4$ CT we  have to insert more $\cH$-covariant space-time derivatives in \rf{la}, so that  the operator $\Delta$ in \rf{ct} is more general than in eq. \rf{DiffOp}. 
 Our main result for the deformed action in \rf{Faction} will depend only on the operator $\Delta$ not transforming under duality.
This holds regardless of the number of derivatives it contains, since each of them is  inert under duality transformations. 
From now on  one should understand the operator $\Delta$ in \rf{ct} as representing $D^{2k}R^4$ CT.

Since the dual field strength $G$ is defined in terms of the field strength $F$ through \eqref{gtensor}, to carry out perturbative calculation it is necessary, on the one hand, to express 
explicitly $G$ in terms of $F$. On the other, adding a deformation such as \rf{la}, depending on both $F$ and $G$, to the classical action defines the action implicitly, by relating it to its 
partial derivatives with respect to $F$. 
Thus, to carry out perturbative calculations with the deformed action it is necessary to solve this differential equation; the solution will generically exhibit arbitrarily-high powers of the 
deformation parameter $\lambda$. 
An alternative approach, which we will carry our in the next section and Appendix~\ref{solve_twisted_SD}, is to determine $G$ by solving a deformed twisted self-duality constraint. 
The deformation of the classical twisted self-duality constraint is chosen such that the leading (i.e. ${\cal O}(\lambda)$) term reproduces the CT deformation of the classical action. 
There are many such deformations of the classical twisted self-duality constraint, which differ by terms of order ${\cal O}(\lambda^{n\ge 2})$. In the discussion in the next section 
and Appendix~\ref{solve_twisted_SD} we shall assume that no such higher-order terms are present.

Adding more derivatives, corresponding to superpartners of $D^{2k} R^4$, will not change the general structure of the two-vector vector (and hence its duality properties), but 
will change the dimension of the CT and the number of loops were it might be generated.  In the context of the four-graviton amplitude it corresponds to  an insertion of a 
dimension-increasing function of Mandelstam variables $f(s,t,u)$. The operator $\Delta$ in such case will have additional  derivatives compared with the expression shown in \rf{DiffOp}.

\section{ Complete two-vector deformed action with duality symmetry}

The {\it twisted nonlinear selfduality constraint} in classical supergravity at $\lambda=0$ was proposed in \cite{Cremmer:1979up}, \cite{deWit:1982bul}.
In $\cH$-covariant form  it states that there are only $n_v$ physical vectors. The constraint is
 \be
 T^+_{\mu\nu\, AB} =  h_{\Lambda
 AB}\, F^{+\Lambda}_{\mu\nu} - f^\Lambda{}_{AB} \,G_{ \mu\nu\,\Lambda}^+  =0  \,,\label{tiden0H}
\ee
together with its complex conjugate.
If instead of using the $\cH$-covariant constraint we would like to use the $\cG$-covariant one, we can multiply the equation on $f^{-1}$ so that
\be
  G_{ \mu\nu\,\Lambda}^+ - (f^{-1}h)_{\Lambda \Sigma
}\, F^{+\Sigma}_{\mu\nu }  =0 \, \qquad \Rightarrow \qquad G_{ \mu\nu\,\Lambda}^+ - \cN_{\Lambda \Sigma
}\, F^{+\Sigma}_{\mu\nu }  =0
 \,.\label{Gform}
\ee

A non-vanishing deformation on the right-hand side of these equations, which  would also be Lorentz and $\cH$-covariant, was presented in eq. (5.7) in 
 \cite{Carrasco:2011jv}.  It can be derived, following the proposal in \cite{Bossard:2011ij} to use the manifestly duality invariant source of deformation. 
 In this case it depends on a duality doublet  ${\cal F}= (F, G)$; that is, the classical twisted self-duality constraint \rf{Gform} is not valid and we propose that 
 its right-hand side is given by the source of deformation
\be
{\cal I}=  \lambda \,   T^{- }_{ AB} \,  \Delta \, \bar  T^{- \,  AB}   = \lambda (  h_{\Sigma
 AB} F^{- \Sigma} - f^\Sigma{}_{AB} \, G_\Sigma^- ) \Delta (\bar h_{\Lambda}{}^{
 AB}\,  F^{+ \Lambda} - \bar f^ {\Lambda AB} \, G_{\Lambda}^+)  \ .
\ee 
It leads to a constraint of the type given in eq. (5.7) in  \cite{Carrasco:2011jv}
\be
T^+_{ AB} + \lambda \Delta T^-_{ AB} =0
\label{twistedCon}\ee
where the $\cH$ covariant differential operator $\Delta$ is defined in eq. \rf{DiffOp}. In fact all results below are valid in a more general case when $\Delta$ depends also on scalars and gravitons.
For the subsequent analysis it is convenient to switch to a $\cG$-covariant form of equations
\be
(f^{-1})^{AB}{}_\Lambda \Big ( T^+_{ AB} + \lambda \Delta T^-_{ AB} \Big )= 0 \; , \label{newG} 
\ee
which will give us the following (we skip indices, they are easy to restore)
\be
 [ G^+ - {\cal N}\, F^{+}  + X ( G^- - \cN F^-) ]_\Lambda =0 \,,\label{Gplus1}
\ee
and the complex conjugate is
\be
 [ G^- - \bar {\cal N}\, F^{-}  + \bar X ( G^+ - \bar \cN F^+) ]_\Lambda =0 \,.
 \label{Gminus1}
\ee
Here the differential operators $X$ and $\bar X$ are 
\be
X=\lambda  f^{-1} \Delta f, \qquad \bar X= \lambda \bar f ^{-1} \bar \Delta \bar f \ .
\ee
We may substitute $G^-$ from \rf{Gminus1} into \rf{Gplus1} and we get
\begin{align}
G^+_\Lambda = \Big [ (1-X\bar{X})^{-1}[X(\cN-\bar{\cN})F^-+(\cN-X\bar{X}\bar{\cN})F^+]\Big] _\Lambda \ .
\label{Gplus}
\end{align}
This can be integrated to produce the deformed action, so that the derivative of the action over $F^+$ will produce the value of $G^+$ in \rf{Gplus}.
The result is
\begin{align}
\mathcal{L}_{\rm def}=-i F^+(1-X\bar{X})^{-1}X(\cN-\bar{\cN})F^- - i F^+(1-X\bar{X})^{-1}({\cN}-X\bar{X}\bar{\cN})F^++{\rm h.c.} \ .
\label{Faction}
\end{align}
The integrability condition requires that 
\be
{\delta G^+_\Lambda \over \delta F^{+ \Sigma}}={i\over 2} {\delta^2 S\over \delta F^{+ \Lambda} \delta F^{+ \Sigma}}\, , 
 \qquad {\delta G^+_\Lambda \over \delta F^{- \Sigma}} ={i\over 2}{\delta^2 S\over \delta F^{+ \Lambda} \delta F^{- \Sigma}}= 
-{\delta G^-_\Sigma \over \delta F^{+ \Sigma}} \ .
\label{int1} \ee
We  test the integrability condition in the appendix B and show that the action \rf{Faction} leads to \rf{Gplus}. And since every term in the expression for $G$ is linear in $F$, it is easy to present a nice and simple form of the vector-dependent part of the action, it is given in the form
\be
\mathcal{L}_{\rm def}= F \tilde {G}.
\label{GZaction}\ee
In conclusion of this section, we have derived a deformed action \rf{Faction}, \rf{GZaction} for $\cN\geq 5$ supergravity, with terms with higher derivatives of an infinite order, which has a duality current conservation. 
It extends the results of \cite{Bossard:2011ij} by giving a closed form expression of the duality-invariant two-vector part of 
the allowed ${\cal N}\ge 5$ counterterm.
The first deformation term, is proportional to $X=\lambda  f^{-1} \Delta f$ and has 8 derivatives, other terms with $X^{n}$ are of the order $\lambda^n \partial^{2n}$. Since now $G^+= {i\over 2} {\delta \cL_{\rm def}\over \delta F^+}$, we find that deformed equations of motion for the $F$-field become exact Bianchi identity for the $G$-field.

\section{ Duality  restoration in an example: $\lambda^2$ approximation, no scalars \label{exampleLaSq}}

Our deformed (bosonic) action is given in eq. \rf{Faction}. We are interested in vector-dependent terms which are independent, 
linear and quadratic in $X\sim \lambda$ 
\begin{align}
\mathcal{L}^{0+1+2}=& - i F^+{\cN}F^+ -i F^+X(\cN-\bar{\cN})F^-  - i F^+X\bar{X}({\cN}-\bar{\cN})F^+
\cr
&+i F^-\bar {\cN}F^- +i F^+ ( \bar{\cN} -\cN) \bar X F^- +i F^-(\bar{\cN} -{\cN}) \bar X {X}F^- \ .
\label{action}
\end{align}
The action has a manifest $SO(\cN)$ symmetry.
We stress that restoration of the $SU({\cal N})$ symmetry for the S-matrix following from this action is a necessary {\em but in general not sufficient} condition for consistency of eq.~\eqref{action} 
and supersymmetry; it is this necessary condition that we shall verify below.

At the base point of the coset space we will take $\cN= -{i}$, $\cN-\bar{\cN}=-2i$, $f=1/\sqrt 2 $, $X=\lambda f^{-1}\Delta f= \lambda \Delta$ and we take $\Delta = \Delta^\dagger$
\begin{align}
\mathcal{L}_{\rm base}^{0+1+2}=- \Big [( F^+)^2  + (F^-)^2\Big]
-  4\lambda F^+\Delta  F^-  -  2\lambda^2 F^+\Delta ^2 F^+ -  2\lambda^2 F^-\Delta ^2 F^- \ .
\label{action0}
\end{align}
In such case we defined the dual field strength as
\be
\tilde { G }= \frac12 {\delta \cL\over \delta F} \ .
\ee

To check directly that the current conservation, broken due to terms $\lambda$ \cite{Kallosh:2011dp}, \cite{Kallosh:2011qt} and restored by the terms of order  $\lambda^2$ in the action 
 we need to compute the $B$ component of the duality current conservation $\partial _\mu J^{\mu  \Lambda \Sigma} B_{\Lambda \Sigma}$.
 The $B$ component of the Gaillard-Zumino duality current $J^\mu_{\rm GZ}B= \tilde G^{\mu\nu}   B   {\cal B}_\nu$, corresponding to the transformation $F^{ \Lambda '} = A^\Lambda{}_\Sigma F^{ \Sigma } + B^{\Lambda \Sigma} G_\Sigma$, can be defined only in the presence of the equation of motion $dG=0$, i.e. in the presence of the dual vector ${\cal B}_\nu$ 
 such that $G=d{\cal B}$.

Here we will just check that, in absence of scalars, $\partial _\mu J^{\mu  \Lambda \Sigma} B_{\Lambda \Sigma}$ vanishes through ${\cal O}(\lambda^2)$. 

This component of the duality current\footnote{ The position of duality indices $\Lambda$ was not specified strictly 
at the level  of \cite{Gaillard:1981rj}-\cite{Kallosh:2008ic}, as it becomes later when in symplectic sections upper component 
was taken with the duality index up, and lower component with the index down.} 
\be
\partial _\mu J^{\mu  \Lambda \Sigma} B_{\Lambda \Sigma} = - \Big ({\delta S\over \delta F^{+ \Lambda }} {\delta S\over \delta F^{+ \Sigma }} - {\delta S\over \delta F^{- \Lambda }} {\delta S\over \delta F^{- \Sigma }}\Big) B_{\Lambda \Sigma}= G^+_{ \Lambda } B^{\Lambda \Sigma} G^+_{ \Sigma } - G^-_{ \Lambda } B^{\Lambda \Sigma} G^-_{ \Sigma } 
= 2 i G_\Lambda B^{\Lambda\Sigma}  \tilde G_\Sigma \ ,
\label{JcomponentB}
\ee
with the self-dual and anti-self-dual dual field strengths given by
\be
-{\delta S_{\rm base}^{0+1+2}\over \delta F^{+ \Lambda }}=2 F^{+\Lambda} +4 \lambda \Delta \cF^{-\Lambda}  + 4\lambda^2 \Delta ^2 \cF^{+\Lambda},
\ee
\be
-{\delta S_{\rm base}^{0+1+2}\over \delta F^{- \Lambda }}= 2F^{-\Lambda} + 4\lambda \Delta F^{+\Lambda} +4  \lambda^2 \Delta ^2 F^{-\Lambda} \ .
\label{BFM}\ee
Eq.~\eqref{JcomponentB} becomes then
\bea
G^+_{ \Lambda } G^+_{ \Sigma } - G^-_{ \Lambda }  G^-_{ \Sigma }
&=&
       (F^+_{ \Lambda }  + 2\lambda \Delta F^-_{ \Lambda }  +  2\lambda^2 \Delta ^2 F^+_{ \Lambda }  )
       (F^+_{ \Sigma } + 2\lambda \Delta F^-_{ \Sigma } +  2\lambda^2 \Delta ^2 F^+_{ \Sigma } )
\cr
&-&  (F^-_{ \Lambda } + 2\lambda \Delta F^+_{ \Lambda } +  2\lambda^2 \Delta ^2 F^-_{ \Lambda } )
        (F^-_{ \Sigma } + 2\lambda \Delta F^+_{ \Sigma } +  2\lambda^2 \Delta ^2 F^-_{ \Sigma } )
\eea
which, up to terms of order ${\cal O}(\lambda^3)$ is a total divergence
\be
G^+_{ \Lambda } G^+_{ \Sigma } - G^-_{ \Lambda }  G^-_{ \Sigma } = F^+_{ \Lambda } F^+_{ \Sigma } - F^-_{ \Lambda } F^-_{ \Sigma } + {\cal O}(\lambda^3) \ . 
\ee
This supports and illustrates at the $\lambda^2$ level the general proof in section~\ref{TDG} that our deformed action has a duality current conservation.

\section{$SU(\cN)$ restoration from $SO(\cN)$ in the six-point amplitude example}

\begin{figure}[t]
\begin{center}
\includegraphics[width=0.6\textwidth]{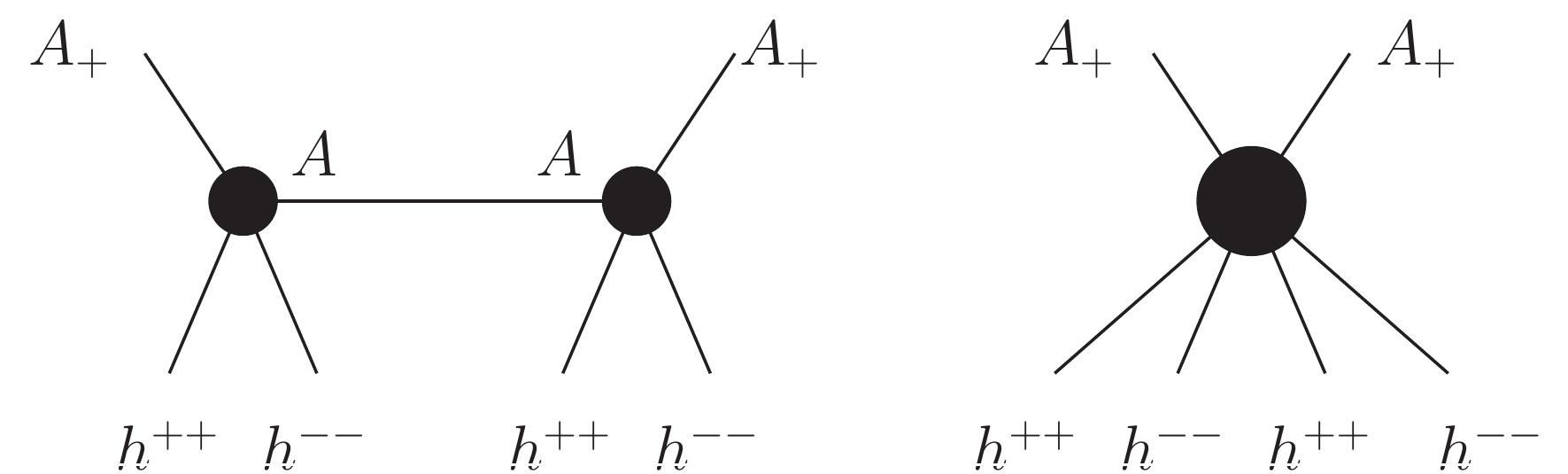}
\end{center}
\caption{\small Graphs contributing to the amplitude ${\cal A}(A^{AB}_+(p_1)A^{CD}_+(p_2)h_{++}(p_3)h_{++}(p_4)h_{--}(p_5)h_{--}(p_6))$.}
\label{fig6pt}
\end{figure}

Note that the deformed action in \rf{action0} has terms with $SO(\cN)$ symmetry, for example using indices we have  $\lambda^2 (F^{+ AB})\Delta^2 (F^{+ AB})$ as well as terms with $SU(\cN)$ symmetry, like $\lambda F^{+ AB}\Delta F^-_{AB}$. In classical theory there are also  $SO(\cN)$ invariant terms, like $ (F^{+ AB})^2$, however, the on shell action is known to have an $SU(\cN)$ symmetry. Here we will find out if the presence of the new $SU(\cN)$ symmetry breaking terms, like $\lambda^2 (F^{+ AB})\Delta^2 (F^{+ AB})$, affects the on shell symmetry of the theory. For this purpose we will compute all contributions to the  $\lambda^2$ amplitude, the one from the single $\lambda^2 (F^{+ AB})\Delta^2 (F^{+ AB})$ vertex and the one from the tree diagram with two vertices  $\lambda F^{+ AB}\Delta F^-_{AB}$, as shown in Figure~\ref{fig6pt}.

In this section we will treat the parameter $\lambda$ as independent of the gravitational coupling.
To test the on-shell symmetry properties of the deformed action \eqref{action0} it therefore
suffices  to analyze tree-level amplitudes with $\lambda$-dependent vertices. Our strategy
will thus be to concentrate on the simplest possible non-trivial tree amplitude involving 
the correction term in lowest order, with four gravitons and two vectors on the external legs,
which is such that no other (of the infinitely many) higher order vertices can contribute. 
To this aim we start from (\ref{action0}) where all dependence on the scalar fields has 
been stripped off. Introducing the chiral projectors
\be
P_\pm{}^{\mu_1\nu_1}_{\rho_1\sigma_1} = \frac{1}{4}\big(\delta_{\rho_1}^{\mu_1}\delta_{\sigma_1}^{\nu_1} - \delta_{\rho_1}^{\nu_1}\delta_{\sigma_1}^{\mu_1} \mp i \epsilon^{\mu_1\nu_1}{}_{\rho_1\sigma_1}\big) 
\ee
onto the self-dual and anti-self-dual components of 2-forms we can schematically
represent the operator $\Delta$ in the form
\begin{eqnarray}
\Delta &=& P_+ (Xhh) P_-  \,+\,  P_- (Xhh) P_+ + {\cal O}(h^3)\,, \nonumber\\[2mm]
\Delta^2 &=& P_+ (Xhh) P_- (Xhh) P_+   \,+\,  P_- (Xhh) P_+ (Xhh) P_- + {\cal O}(h^5) \,,
\end{eqnarray}
where $Xhh$ is the leading term in the expansion of the fourth order
differential operator $\Delta$ to lowest (quadratic) order in the metric fluctuations.
With this the action (\ref{action0}) contains the following pieces up to 
and including second order in $\lambda$ (still omitting internal indices)
\begin{eqnarray}\label{FDeltaF}
- 4\lambda F^+ \Delta F^-  &=& 
-4\lambda (\partial A) P_+ (Xhh) P_- (\partial A) + {\cal O}(h^3)\,,
\nonumber\\[2mm]
- 2\lambda^2  F^+ \Delta^2 F^+  &=&
- 2\lambda^2 (\partial A) P_+ (Xhh) P_- (Xhh) P_{+} (\partial A) + {\cal O}(h^5) \,,
\nonumber\\[2mm]
- 2\lambda^2 F^- \Delta^2 F^-  &=&
- 2\lambda^2 (\partial A) P_- (Xhh) P_+ (Xhh)  P_- (\partial A) + {\cal O}(h^5)\,.
\end{eqnarray}
 For the computation of the scattering amplitude we must saturate these vertices
with the polarization states $\epsilon^{\pm\pm}_{\mu\nu}(p)$ for the gravitons, and
$\epsilon^\pm_\mu(p)$ for the vectors (with the usual on-shell conditions $p^2 =0$ and
$p^\mu \epsilon^{\pm\pm}_{\mu\nu}(p)  = p^\mu \epsilon^\pm_\mu(p) =0$). Putting
back the internal indices we recall that the vector fields of $\cN$-extended supergravity
transform in the adjoint of $SO(\cN)$ (with an extra singlet vector for $\cN=6$); the vector 
polarizations therefore carry an extra $SO(\cN)$ label $[AB]$. When applied to the field 
strength this $SO(\cN)$ label becomes elevated to an $(S)U(\cN)$ index pair, where we 
must now distinguish between upper and lower positions of the indices $[AB]$. 
For instance, for $\cN=8$ supergravity this results in the substitutions
\ba
F^{+AB}_{\mu\nu} &\rightarrow& i p_{[\mu} \epsilon^{+AB}_{\nu]}  \nonumber\\[2mm]
F^-_{\mu\nu AB}  &\rightarrow& i p_{[\mu} \epsilon^-_{\nu]}{}_{AB}
\ea
where $F^{+AB}_{\mu\nu}$ transforms in the $\bf{28}$ of $SU(8)$, while
$F^{-}_{\mu\nu AB}$ transforms in the $\bf{\overline{28}}$ of $SU(8)$, with 
independent polarizations $\epsilon^{\pm AB}_\mu$ for all  vectors. These $SU(8)$ assignments are furthermore consistent with the relations
\ba\label{Pepsilon}
P_{+\mu\nu}{}^{\rho\sigma} (ip_\rho \epsilon_\sigma^{+AB})  \,&=&\,
      ip_{[\mu} \epsilon_{\nu]}^{+AB} \;\;\;, \quad
P_{-\mu\nu}{}^{\rho\sigma} (ip_\rho \epsilon_\sigma^{-}{}_{AB})  \,=\,  
      ip_{[\mu} \epsilon_{\nu]}^{-}{}_{AB} \,, \nonumber\\[2mm]
P_{+\mu\nu}{}^{\rho\sigma} (ip_\rho \epsilon_\sigma^{-}{}_{AB})  \,&=&\, 0 
 \, \qquad\qquad , \quad
P_{-\mu\nu}{}^{\rho\sigma} (ip_\rho \epsilon_\sigma^{+AB}) \, = \, 0 \,.
\ea
At order $\cO(\lambda^2)$ the amplitude
\be
{\cal{A}}^{AB,CD} (p_1, \dots, p_6) = 
\Big\langle A^{AB}_+(p_1)A^{CD}_+(p_2)h_{++}(p_3)h_{++}(p_4)h_{--}(p_5)h_{--}(p_6)\Big\rangle
\ee
will thus receive two contributions, namely one from the square of the quadratic vertex (first line
in (\ref{FDeltaF})) with two vectors contracted, and the other from the sextic vertex (second and 
third line in (\ref{FDeltaF})); these two contributions are depicted in Figure~\ref{fig6pt}. We note 
that in this amplitude both index pairs $[AB]$ and $[CD]$ 
are in the upper position because of the positive helicities 
of the external spin-one states. Since one cannot form an $SU(\cN)$ singlet with four  upper 
indices for $\cN\geq 5$, a non-vanishing result for this amplitude would indicate a breakdown 
of the $SU(\cN)$ R symmetry. 
However, we will now show that this amplitude indeed vanishes.

To proceed we first consider the square of the $\cO(\lambda)$ vertex: not forgetting a
factor $1/2$ from the expansion of the exponential this leads to
\ba\label{squarevertex}
&& \frac12 \big( -4i\lambda (\partial A) P_+ (Xhh) P_- (\partial \underbracket{A)\big)
\big( -4i\lambda (\partial A}) P_- (Xhh) P_+ (\partial A) \big)\, %+ \nonumber \\[2mm]
\ea
where the underbracket denotes the contraction (= vector propagator in a convenient gauge)
\be
\underbracket{A_\mu^{AB}(k) A_\nu^{CD}}(-k) 
\,=\,  - \frac{i}{k^2} \, \eta_{\mu\nu} \delta^{B[A} \delta^{C]D}
\ee
and where the positive helicity vectors are left uncontracted as they will be dressed
with positive helicity polarizations in accord with (\ref{Pepsilon}). Now using the relation
\be
P_-{}^{\mu_2\nu_2}_{\rho_2\sigma_2} k{}_{\mu_2} \eta_{\nu_2 {\bar\nu}_2}P_-{}^{{\bar\mu}_2{\bar\nu}_2}_{{\bar\rho}_2{\bar\sigma}_2} k{}_{{\bar\mu}_2}
=\frac{1}{4}k^2 P_-{}_{\rho_2\sigma_2;{\bar\rho}_2{\bar\sigma}_2}, 
\ee
with the momentum $k= p_1 + p_3 + p_5$  (= $-p_2 - p_4 - p_6)$ on the internal line
we see that the propagator factor is cancelled, and we end up with an effective
{\em local} vertex
$$
+ \, 2i\lambda^2 (\partial A) P_+ (Xhh) P_- (Xhh) P_{+} (\partial A)
$$
which is the same as the contact interaction in the second line in (\ref{FDeltaF}). Therefore the two contributions exactly cancel at the order of $\lambda^2$. Thus, the deformed all order higher derivatives action,  which has a duality current conservation, yields a six-point on-shell amplitudes 
at the $\lambda^2$ order which does  exhibit the expected $SU(\cN)$ demanded by $\cN$-extended supergravity. Using the vertices in eq \rf{FDeltaF} it is not difficult to show that 
the eight-point ${\cal O}(\lambda^3)$ $SU(\cN)$-breaking amplitude also vanishes.

To conclude, we have shown  that the bosonic duality-symmetric action with higher derivatives does not break the $SU(\cN)$ symmetry of the six- and eight-point on shell amplitudes to $SO(\cN)$. However, in general, the issue of the restoration of $SU(\cN)$  symmetry and supersymmetry based on deformed action requires additional investigations.

\section{Discussion}
In this paper we have constructed a complete deformed action of the two-vector sector of the candidate UV divergence serving as the seed of deformation of $\cN\geq 5$ supergravities; the resulting action terms in eqs.~\eqref{Gplus} and \eqref{Faction} extend earlier results of \cite{Bossard:2011ij}, where the duality-completion of the two-vector superpartner of an $R^4$ counterterm was first considered. We have solved perturbatively the twisted non-linear constraint equation \rf{twistedCon} and identified the dual field strength $ G^+(F, \phi) $ to all orders in $\lambda$, presented in \rf{Gplus}. We have also found the complete all order in $\lambda$ action \rf{Faction} such that the corresponding duality  current is conserved.

Our deformed action, when expanded near the base point of the moduli space ${\cG\over \cH}$ has terms which break 
$SU(\cN)$ symmetry down to $SO(\cN)$ symmetry. This feature, if it would persist on shell, would prevent our deformed action from being consistent with supersymmetry. We have therefore computed the six-point amplitude, as shown in Figure~\ref{fig6pt}, and we have found that the contribution from the  $SU(\cN)$ symmetry violating $\lambda^2$ vertex in the deformed action is precisely cancelled by the tree diagram with two $\lambda$ vertices. These examples  indicate that an analogous cancellation and restoration of $SU(\cN)$ symmetry in scattering amplitudes might take place at all higher orders in $\lambda$ and for all $n$-point amplitudes.

Our conclusion here is the following.  When using  the two-vector sector of the candidate counterterm as a seed for deformation of the action we do not  find an inconsistency between the requirement of duality current conservation and supersymmetry of the deformed action. It does not mean that our deformed action has a supersymmetric embedding, but there is also no obvious obstruction to it: the six-point tree amplitude based on deformed action has an $SU(\cN)$ symmetry, which is necessary but not sufficient condition for supersymmetry. 

Our analysis here does not explain why $\cN=5$ supergravity in four loops is UV finite~\cite{Bern:2014sna}. We will continue with analogous investigation of  more general sectors of the deformation of the theory in Part II of this project. We will take into account the one-vector and the four-vector sectors, in addition to the two-vector sector we have studied here. Ultimately, the goal is to  either construct a supersymmetric deformed action of $\cN\geq 5$ supergravity, or to find that it is not available.

\vspace*{1cm}

 \noindent{\bf {Acknowledgments:}} We are grateful to Z. Bern,  L. Brink, J.J. Carrasco, D. Freedman, M. Green, M. Duff, S. Ferrara, H. Elvang, Y.-T. Huang, P. Townsend, A. Tseytlin and  A. Van Proeyen  for stimulating discussions.  The work  of  RK and YY is supported by SITP and by the US National Science Foundation grant PHY-1720397. 
 The work of HN is suppoted in part by the European Research Council under the European Union's Horizon 2020 research and innovation
   program (grant agreement No 740209). The work of RR is supported in part by the US Department of Energy under grant DE-SC0013699.
 RK and HN are grateful to the organizers of SUSY 2016 conference in Melbourne where this work was initiated.
 RR is grateful to SITP at Stanford and RK to Max Planck Institute of Gravitational Physics in Potsdam for the hospitality when a part of this work was performed.
 We are grateful to the participants of the conference `Hawking 75' in Cambridge for the interest to our work and important discussions.

\appendix 
\section{   A review of classical $\cN$-extended supergravities}
 \label{gen4d} 
 
 We start by recalling \footnote{This is a shortened version of the corresponding  review in \cite{Andrianopoli:2006ub}, which focuses on the details important to our case.} the 
 main features of four dimensional pure $\cN$-extended supergravities,  $\cN\geq 5$.

These theories contain in the bosonic sector,
besides the metric, a number $n_v$ of vectors and $m$ of (real)
scalar fields. The relevant classical bosonic vector and scalar part of action is known to have the
following general form:
\begin{eqnarray}
 {\cal S}&=&\int\sqrt{-g}\,
d^4x\left(-\frac{1}{2}\,R+{ \rm Im} {\cal N}_{\Lambda
\Gamma}F_{\mu\nu}^{\Lambda } F^{\Gamma |\mu\nu}+
\frac{1}{2\,\sqrt{-g}}\,{\rm Re}{\cal N}_{\Lambda \Gamma  }
\epsilon^{\mu\nu\rho\sigma}\, F_{\mu\nu}^{\Lambda } F^{\Gamma
}_{\rho\sigma}+\right.\nonumber\\
&+&\left.\frac 12 \,g_{rs}(\Phi) \partial_{\mu}
\Phi^{r}\partial^{\mu}\Phi^{s}\right)\,.\label{bosonicL}
\end{eqnarray}
The vector-scalar part of this action was presented in \rf{lagrapm} and notations explained there.
Duality rotations and symplectic covariance of these theories were uncovered in \cite{Gaillard:1981rj}.
\par
We consider a theory of  $n_v$ abelian gauge fields
$A^\Lambda_{\mu}$, in a $D=4$  space-time with Lorentz signature
(which we take to be mostly minus). They correspond to a set of
$n_v$ differential $1$-forms
\begin{equation}
A^\Lambda ~ \equiv ~ A^\Lambda_{\mu} \, dx^{\mu} \quad \quad
 \left ( \Lambda = 1,
\dots , {n_v} \right )\,.
\end{equation}
The corresponding field strengths and their Hodge duals are
defined by \footnote{We use, for the $\epsilon$ tensor, the convention: $\epsilon_{0123}=-1$.}
\begin{eqnarray}
{ F}^\Lambda & \equiv & d \, A^\Lambda \,  \equiv   \,
    F^\Lambda_{\mu  \nu} \,
dx^{\mu } \, \wedge \, dx^{\nu} ,\nonumber\\
 F^\Lambda_{\mu \nu} & \equiv & {\frac{1}{2 }} \,\left (
\partial_{\mu } A^\Lambda_{\nu} \, - \, \partial_{\nu }
A^\Lambda_{\mu} \right ), \nonumber\\ ({}^{\star}
{F}^{\Lambda})_{\mu\nu} & \equiv &\, {\frac{\sqrt{-g}}{2}}
\varepsilon_{\mu \nu \rho\sigma}\,  F^{\Lambda \vert \rho \sigma}\,.
\label{campfort}
\end{eqnarray}
The dynamics of a system of abelian gauge fields coupled to scalars in a
 gravity theory is encoded in the bosonic action
(\ref{bosonicL}).
Introducing self-dual and
anti-self-dual combinations
\begin{eqnarray}
   F^{\pm} &=& {\frac 12}\left( F\, \pm {\rm i} \,
 {}^ \star F\right)
 %\,, \nonumber \\
 \quad , \qquad 
{}^\star (F^{\pm}) = \mp \mbox{i}  F^{\pm} \  , 
\label{selfduals}
\end{eqnarray}
the vector part of the Lagrangian defined by (\ref{bosonicL}) can
be rewritten in the form given in \rf{lagrapm}
We introduce  new tensors
\begin{equation}
  {{}^\star G}_{\Lambda|\mu\nu} \, \equiv \,  {\frac 12 } { \frac{\partial
{\cal L}}{\partial  F^\Lambda_{\mu\nu}}}= { \rm Im} {\cal
N}_{\Lambda\Sigma}\,F^\Sigma_{\mu\nu}  +{ \rm Re} {\cal
N}_{\Lambda\Sigma}\, {}^\star F^\Sigma_{\mu\nu} 
~~\longleftrightarrow~~
G^{\mp }_{\Lambda |\mu\nu} \, \equiv \, \mp{\frac {\rm i}{2} } {
\frac{\partial {\cal L}}{\partial  F^{\mp \Lambda}_{\mu\nu}}}\,,
\label{gtensor}
\end{equation}
the Bianchi identities and field equations associated with the
Lagrangian (\ref{bosonicL}) can be written as 
\bea
\nabla^{\mu }{{}^\star  F}^{\Lambda}_{\mu\nu} &=& 0 
\quad , \qquad
\nabla^{\mu }{{}^\star G}_{\Lambda |\mu\nu} = 0 
\label{biafieq}
\eea 
or equivalently 
\bea
\nabla^{\mu }{\rm Im} F^{\pm \Lambda}_{\mu\nu} &=& 0 
\quad, \qquad
\nabla^{\mu } {\rm Im} G^{\pm }_{\Lambda |\mu\nu} = 0 \ .
\label{biafieqpm} 
\eea 
Introducing the $2{n_v}$-component column vector
\begin{equation}
{{}^\star \mathcal{F}}  \, \equiv \, \left(
  \begin{array}{c}
  {{}^\star F}^{\Lambda} \cr  {{}^\star G}_{\Lambda} \cr  
  \end{array}
\right) \ ,
\label{sympvec}
\end{equation} 
a general duality rotation is any general linear transformations on such a vector,
\begin{equation}
\left ( \begin{array}{c} {}^{\star}  F\cr {}^{\star}  G\cr   \end{array}\right
)^\prime \, =\, \left ( \begin{array}{cc} A & B \cr C & D \cr  \end{array} \right )
\left (  \begin{array}{c} {}^{\star}  F\cr {}^{\star}  G\cr  \end{array}\right ).
\label{dualrot}
\end{equation}
For any constant matrix ${\cal S} =\left (\begin{array} {cc}A & B \cr C & D
\cr  \end{array} \right ) \, \in \, GL(2{ n_v},\mathbb{R} )$ the transformed vector of
magnetic and electric field-strengths ${{}^\star \mathcal{F}}^\prime ={\cal S}
\cdot {{}^\star \mathcal{F}}$ satisfies the same equations ~\rf{biafieq} as the
original one. 
In a condensed notation we can write
\begin{equation}
\partial \, {{}^\star \mathcal{F}}\, = \, 0 \quad \Longleftrightarrow \quad
\partial \, {{}^\star \mathcal{F}}^\prime \, = \, 0.
\label{dualdue}
\end{equation}
Separating the self-dual and anti-self-dual parts
\begin{equation}
 F = F^+ + F^-  \quad ; \qquad
 G = G^+ + G^-  
\label{divorzio}
\end{equation}
and taking into account that  $F$ and $G$ are related by \eqref{gtensor}, 
\begin{equation}
 G^+ \, = \, {\cal N} F^+  \qquad ; \qquad  G^- \, = \, {\bar {\cal N}} F^-  \ ,
\label{gigiuno}
\end{equation}
the duality rotation in eq.~\rf{dualrot} can be rewritten as
\begin{equation}
\left (  \begin{array}{c}   F^+ \cr
  G^+ \cr  \end{array}\right )^\prime  \, = \,
{\cal S} 
\left (  \begin{array}{c} F^+\cr {\cal N}  F^+\cr  \end{array}\right ) \qquad ; \qquad \left (  \begin{array}{c}
  F^- \cr
  G^-\cr  \end{array}\right )^\prime \, = \,
{\cal S} \left (  \begin{array}{c}
F^-\cr {\bar {\cal N}}  F^-\cr   \end{array}\right ) . \label{trasform}
\end{equation}
The kinetic matrix $\N = \cN(\Phi)$
transforms under a duality rotation such that the definition of $G^\mp$ as a variation of the Lagrangian continues to hold:
\begin{equation}
G^{\prime +}_{\Lambda} = \left(C +
D \cN\right)_{\Lambda\Sigma} F^{+\Sigma} \equiv - \frac{\rm i
}{ 2} \frac{\partial {\cal L}^\prime}{  \partial  F^{\prime +
\Lambda}}= \left(A + B \cN\right)^\Delta_{\ \Sigma}
 \cN^\prime_{\Lambda\Delta} F^{+\Sigma}
\end{equation}
that
\begin{equation}
\cN^\prime_{\Lambda\Sigma}(\Phi^\prime) = \left[\left(C +
D \cN\right) \cdot \left(A +
B \cN\right)^{-1}\right]_{\Lambda\Sigma} . \label{ntrasfeven}
\end{equation}
The condition that the matrix $\cN$ is symmetric both before and after the duality transformation  implies that
  \begin{equation}  {\cal S}\in Sp(2n_v, \mathbb{R})\subset GL(2n_v, \mathbb{R})\,,
  \end{equation}
  that is:
\begin{eqnarray}
{\cal S}^T\,\mathbb{C}\,{\cal S}&=&\mathbb{C}\,,\label{syms}
\end{eqnarray}
where $\mathbb{C}$ is the symplectic invariant $2n_v\times 2 n_v$
matrix:
\begin{eqnarray}
\mathbb{C}&=&\left(\begin{array}{cc} 0 & -\bfone\cr \bfone &
0 \end{array}\right)\,.\label{C}
\end{eqnarray}
It is useful to rewrite the symplectic condition \rf{syms} in
terms of the $n_v\times n_v$ blocks defining~${\cal S}$:
\begin{eqnarray}
A^T\,C- C^T\,A&=& B^T\,D-D^T\,B=0\,\,\,;\,\,\,\,
A^T\,D-C^T\,B=\bfone\,.\label{abcd}
\end{eqnarray}

In $\cN\geq 5$ models the fields are in some representation of the isometry group ${\cG}$  of the scalar manifold or of its maximal compact subgroup $\cH$. 
\footnote{This group is also the isotropy group of the scalar manifold and it is also isomorphic to the R-symmetry group; we use these names interchangeably  when referring to $\cH$.}
All the properties of supergravity theories for $\cN\geq 5$ are completely fixed in terms of the geometry of the coset ${\cG/\cH}$; they can be formulated in terms of the 
coset representatives $L$ satisfying by
 \begin{equation}
 L(\Phi^\prime) =g L(\Phi) h (g,\Phi) \ .
\end{equation}
Here $g\in {\cG}$, $h\in \cH$  and $\Phi ^\prime =   \Phi ^\prime (\Phi)$, $\Phi$ being the coordinates of ${\cG}/\cH$.
Note that the  scalar fields in ${\cG}/\cH$ can be assigned, in the
linearized theory,  to linear representations
 $R_\cH$ of the local isotropy group  $\cH$ so that dim $R_\cH$ = dim ${\cG}$ $-$ dim $\cH$ 
 (in the full theory, $R_\cH $ is the representation which the vielbein of ${\cG}/\cH$ belongs to).

 Fermions in extended supergravities form representations the isotropy subgroup $\cH$ rather than of the 
isometry group $\cG$ of the scalar manifold. For example,  there is the graviphoton -- 
  2-form $T_{AB}$ -- appearing in the supersymmetry transformation law of the gravitino 1-form
\begin{equation}
\delta \psi_A = \nabla\epsilon_A + \alpha T_ {AB\vert
\mu\nu}\gamma^a\gamma^{ \mu\nu} \epsilon^B V_a+ \cdots.
\label{tragra}
\end{equation}
Here $\nabla$ is the covariant derivative in terms of the
space-time spin connection and the composite connection of  $\cH$, $\alpha$ is a coefficient fixed by
supersymmetry, $V^a$ is the space-time vielbein, $A=1,\cdots ,\cN$
is the index acted on by the automorphism group $\cH$ in the fundamental representation. 
Here and in the following
 the ellipsis denote trilinear fermion terms.
The 2-form field strength $T_{AB}$ will be constructed by dressing
the bare field strengths  $ F^\Lambda$ with the coset representative
$L(\Phi)$ of ${\cG/\cH}$, $\Phi$ denoting a set of coordinates of $\cG/\cH$. 
The same field strength $T_{AB}$ which appears in the gravitino
 transformation law is also present in the dilatino transformation law
  \begin{equation}
\delta \chi_{ABC} = P_{ABCD,\ell}\partial_\mu \phi^\ell \gamma^\mu
\epsilon^D +\beta T_{[AB\vert \mu\nu}\gamma^{\mu\nu} \epsilon_{C]}
+\cdots.   \label{tradil}
\end{equation}
Here $P_{ABCD}= P_{ABCD,\ell}d\phi^\ell$ 
is the vielbein of the  scalar manifold, $\beta$ is a coefficient fixed by
supersymmetry.
 
In order to give the explicit dependence on scalars of $T_{AB}$
 it is necessary to recall  that,
according to the Gaillard--Zumino construction, the isometry group
${G}$ of the scalar manifold acts on the
  vector $( F^{- \Lambda},G^{-}_\Lambda)$
  (or its complex conjugate) as a subgroup of
  $Sp(2 n_v,\mathbb{R})$ ($n_v$ is the number of vector fields)
with duality transformations interchanging electric and magnetic
 field strengths, as shown in \rf{trasform}
 
Let now $L(\Phi)$ be the coset representative of $\cG/\cH$ in the
symplectic representation, namely as a $2\,n_v\times 2\,n_v$ matrix
belonging to  $Sp(2n_v,\mathbb{R})$ and therefore, in each theory,
it can be described in terms of $n_v\times n_v$ blocks
$A_L,B_L,C_L,D_L$ satisfying the same relations (\ref{abcd}) as the
corresponding blocks of the generic symplectic transformation ${\cal S}$.

Since the fermions of supergravity theories transform in a complex
representation of the R-symmetry group $\cH \subset \cG$, it is
useful to introduce a complex basis in the vector space of $Sp(2
n_v,\mathbb{R})$, defined by the action of following unitary
matrix:
\begin{eqnarray}
{\cal A}&=& \frac{1}{\sqrt{2}}\,\left(\begin{array}{cc}\bfone & i\,\bfone \cr
\bfone & -i\,\bfone\cr \end{array}\right)\,,\nonumber
\end{eqnarray}
and to introduce a new matrix ${\bf V}(\Phi)$ obtained by
complexifying the right index of the coset representative
$L(\Phi)$, so as to make its transformation properties under right
action of $\cH$ manifest:
\begin{equation}
{\bf V}(\Phi) = \left(\begin{array}{cc}{\bf f} & \bar {\bf f} \cr {\bf h}
&\bar {\bf h} \cr
 \end{array}\right) =
   L(\Phi) \cA^\dagger\,,
  \label{defu}
\end{equation}
where:
\begin{eqnarray}
  {\bf f}&=&\frac{1}{\sqrt{2}} (A_L-{\rm i} B_L)\,\,;\,\,\,
{\bf h} =\frac{1}{\sqrt{2}} (C_L-{\rm i} D_L)\,. \nonumber
\end{eqnarray}
From the properties of $L(\Phi)$ as a symplectic matrix, it is
easy to derive the following properties for ${\bf V}$:
\begin{eqnarray}
{\bf V}\,\eta\,{\bf V}^\dagger &=&
-i \mathbb{C}\,\,\,;\,\,\,\,\,{\bf V}^\dagger\,\mathbb{C}\,{\bf
V} = i\eta\,,\label{propu}
\end{eqnarray}
where the symplectic invariant matrix $\mathbb{C}$ and $\eta$ are
defined as follows:
\begin{eqnarray}
\mathbb{C}&=&\left(\begin{array}{cc} 0& -\bfone \cr \bfone & 0\cr
\end{array} \right)\,\,;\,\,\,\eta = \left(\begin{array}{cc} \bfone & 0 \cr 0 & -
\bfone \end{array}\right)\,,
\end{eqnarray}
and, as usual, each block is an $n_v\times n_v$ matrix. The above
relations imply on the matrices ${\bf f}$ and ${\bf h}$ the
following properties:
 \ba
{\rm i}({\bf f}^\dagger {\bf h} - {\bf
h}^\dagger {\bf f}) &=& \bfone \cr
 ({\bf f}^t  {\bf h} - {\bf h}^t
{\bf f}) &=& 0 .\label{specdef}
\ea

%%%%%%%%%%%%%%%%%%%%%%%%%
%%%%%%%%%%%%%%%%%%%%%%%%%
%%%%%%%%%%%%%%%%%%%%%%%%%

The $n_v\times n_v$  blocks $ {\bf f},\, {\bf h}$ of ${\bf V}$ acquire the following form
\begin{eqnarray}
  {\bf f}&=& f^\Lambda{}_{AB}  \,, \nonumber\\
{\bf h}&=& h_{\Lambda AB} \,, \label{deffh}
\end{eqnarray}
where $AB$ are indices in the two-index antisymmetric representation of
$H= SU({\cal N}) \times U(1)$ or $SU(8)$ in ${\cal N}=8$ case. Upper
$SU(\cN)$ indices label objects in the complex conjugate
representation of $SU(\cN)$: 
\be
(f^\Lambda{}_{AB})^* = \bar{f}^{\Lambda AB} 
\ee etc.
Thus we have another symplectic section depending on scalars of the theory and transforming as follows
 \begin{equation}
\left(\begin{array}{c} f^\Lambda{}_{AB} \cr h_{\Lambda AB}\cr \end{array}\right)^\prime= {\cal S} \left(\begin{array}{c} f^\Lambda{}_{AB} \cr h_{\Lambda AB}\cr \end{array}\right).
\end{equation}

 The kinetic matrix $\cN$ can be written in terms of the
sub-blocks ${\bf f}$, ${\bf h}$, and turns out to be:
\begin{equation}
\cN= {\bf h}\,{\bf f}^{-1}, \quad\quad \cN = \cN^t\,,
\label{nfh-1}
\end{equation}
transforming projectively under $Sp(2n_v ,\mathbb{R})$ duality
rotations as already shown in the previous section. By using
(\ref{specdef})and (\ref{nfh-1}) we find that
   \begin{equation}
   ({\bf f}^t)^{-1} = {\rm i} (\cN - \bar \cN)\bar {\bf f}\,,
\end{equation}
 that is
 \begin{eqnarray}
({\bf f}^{-1})^{AB}{}_{\Lambda} &=&
{\rm i} (\cN - \bar \cN)_{\Lambda\Sigma} \bar{f}^{\Sigma\,AB}\,.
\label{iden}\end{eqnarray}
For the symplectic product in general $\langle ~ \mid ~ \rangle$, one can  use the convention
\begin{equation}
\langle  \mathcal{A}\mid \mathcal{B}\rangle \equiv
  \mathcal{B}^{\Lambda}\mathcal{A}_{\Lambda} 
-\mathcal{B}_{\Lambda}\mathcal{A}^{\Lambda}\, .
\end{equation}
In particular, a symplectic invariant  can be constructed using one symplectic section depending on field strength and its dual $(F^{\pm}, G^{\pm})$ and the other one depending on scalars $(f, h)$
\be
T_{AB}^{\pm}=(f^\Lambda{}_{AB}, h_{\Lambda AB}) \left(\begin{array}{cc} 0& -\bfone \cr \bfone & 0\cr
\end{array} \right)\, \left(\begin{array}{c} F^{\pm \Lambda} \cr G^{\pm}_\Lambda\cr \end{array}\right).
\label{inv}\ee
Here  $T_{AB}^{\pm}$ is a  $\cG$-invariant since 
$
{\cal S}^T\,\mathbb{C}\,{\cal S}=\mathbb{C}$, but it transforms under the group $\cH$. 
Thus, the  graviphoton and its conjugate are
\begin{eqnarray}
T^-_{AB}&=& 
 h_{\Lambda AB}\,  F^{-\Lambda} -f^\Lambda{}_{AB} \,G^-_\Lambda \,, \nonumber\\ 
 \nonumber\\
\bar T^{- AB} &=& (T^-_{AB})^* =\bar h_{\Lambda}{}^{ AB}\,  F^{+\Lambda} -\bar f^{ \Lambda AB} \,G^+_\Lambda\,. \nonumber\\
  \label{gravi}
\end{eqnarray}
Note that, in classical supergravity, the graviphoton satisfies the constraint shown in eq. \rf{tiden0H} as a consequence of eqs. (\ref{nfh-1}), (\ref{gigiuno}). 
It is an $\cH$-covariant form of what is known as a twisted selfduality constraint, covariant under $\cG$ transformations.

 The
constraint eq.~\rf{tiden0H}  is known as \textit{linear twisted
  self-duality constraint}. It can be given in the following form.
We can use a $56$-dimensional real symplectic
vector of field strengths
\begin{equation}
\mathcal{F} \equiv 
\left(
  \begin{array}{c}
  F^{\Lambda} \\ G_{\Lambda} \\  
  \end{array}
\right)\,,
\end{equation}
that transforms in the $\mathbf{56}$ of $E_{7(7)}\subset Sp(56,\mathbb{R})$.
The scalars of the theory are described by the symplectic section
\begin{equation}
\label{eq:symplecticsection}
\mathcal{V}_{AB}
\equiv
\left(
  \begin{array}{c}
  f^{\Lambda}{}_{AB} \\ h_{\Lambda\, AB} \\  
  \end{array}
\right)\, .
\end{equation}
 The period matrix is defined by the
property
\begin{equation}
\label{eq:periodmatrixdef}
h_{\Lambda\, AB} = \mathcal{N}_{\Lambda\Sigma}f^{\Sigma}{}_{AB}\, .  
\end{equation}

This relation of the components of the section $\mathcal{V}_{IJ}$ with the
components of the symplectic $E_{7(7)}/SU(8)$ coset representative imply the
constraints%
\begin{equation}
\langle \mathcal{V}_{AB}\mid\overline{\mathcal{V}}^{\, CD}\rangle 
 =    
-2i\delta_{AB}{}^{CD}\, , 
\hspace{1cm}
\langle \mathcal{V}_{AB}\mid\mathcal{V}_{CD}\rangle =0\, .
\end{equation}

The graviphoton field strength is defined by
\begin{equation}
\label{eq:graviphotondef}
T_{AB}  
\equiv
\langle \mathcal{V}_{AB}\mid \mathcal{F} \rangle\, , 
\end{equation}
and its self- and anti-self-dual parts
are
\begin{equation}
T_{AB}{}^{\pm}  
\equiv
\langle \mathcal{V}_{AB}\mid \mathcal{F}{}^{\pm}   \rangle\, . 
\end{equation}
They all transform under compensating $SU(8)$ transformations only.  Since the
$\cH$-tensor $T_{AB}$ is complex, we have
\begin{equation}
T^{AB\, \pm} = \overline{(T_{AB}{}^{\mp})}\, .  
\end{equation}
Finally, the linear twisted self-duality constraint eq.~(\ref{gigiuno}),
is equivalent to the vanishing of %
\begin{equation}
\label{linear}
\overline{T}^{AB\, -} = \overline{(T_{AB}{}^{+})} =0\, . 
\end{equation}

We are now able to derive some  differential relations  using the Maurer--Cartan equations
obeyed by the scalars through the embedded coset representative
${\bf V}$. Indeed, let $\Gamma = {\bf V}^{-1} d{\bf V}$ be the
$Sp(2n_v,\mathbb{R})$ Lie algebra left invariant one form
satisfying:
\begin{equation}
  d\Gamma +\Gamma \wedge \Gamma = 0\,.
\label{int}
\end{equation}
In terms of $({\bf f},{\bf h})$, $\Gamma$ has the following form:
\begin{equation}
  \label{defgamma}
  \Gamma \equiv {\bf V}^{-1} d{\bf V} =
\left(\begin{array} {cc} {\rm i} ({\bf f}^\dagger d{\bf h} - {\bf h}^\dagger
d{\bf f}) & {\rm i} ({\bf f}^\dagger d\bar {\bf h} - {\bf
h}^\dagger d\bar {\bf f}) \cr -{\rm i} ({\bf f}^t d{\bf h} - {\bf
h}^t d{\bf f}) & -{\rm i}({\bf f}^t d\bar {\bf h} - {\bf h}^t
d\bar {\bf f}) \cr \end{array} \right) \equiv \left(\begin{array} {cc} \Omega^{(H)}&
\bar \cP \cr \cP & \bar \Omega^{(H)} \cr  \end{array}\right)\,,
\end{equation}
where the $n_v \times n_v$ sub-blocks $ \Omega^{(H)}$ and  $\cP$
embed the $\cH$-connection and the vielbein of $\cG/\cH$ respectively.
This identification follows from the Cartan decomposition of the
$Sp(2n_v,\mathbb{R})$ Lie algebra.

From  (\ref{defu}) and (\ref{defgamma}),  we obtain the $(n_v
\times n_v)$ matrix equation:
\begin{eqnarray}
 D(\Omega) {\bf f} &=& \bar {\bf f} \, \cP \,,\nonumber \\
D(\Omega) {\bf h} &=&   \bar {\bf h} \,\cP \,, \label{nablafh}
 \end{eqnarray}
together with their complex conjugates. The $\cH$-connection is
\begin{equation}
\Omega^{(H)} = {\rm i} [ {\bf f}^\dagger (D {\bf h}+ {\bf h}
\omega) - {\bf h}^\dagger (D {\bf f} + {\bf f} \omega)] = \omega
\bfone\,, \label{defomega}
\end{equation}
where we have used:
\begin{equation}
D {\bf h} = \bar \cN D {\bf f} ; \quad {\bf h} = \cN {\bf f}\,,
\end{equation}
which follow from \rf{nablafh} and  the fundamental identity
(\ref{specdef}). Furthermore, using the same relations, the
embedded vielbein  $\cP$ can be written as follows  
\begin{equation}
\cP = - {\rm i} ( {\bf f}^t D {\bf h} - {\bf h}^t D {\bf f}) =
{\rm i} {\bf f}^t (\cN - \bar \cN) D {\bf f} \,,
\end{equation}
 and
\begin{eqnarray}
  D(\omega) f^\Lambda{}_{AB} &=&    \frac{1}{2}\bar f^{\Lambda CD} P_{ABCD}.  \label{df}
\end{eqnarray}
For $N>4$,  $\cP $ coincides with
the vielbein $P_{ABCD}$ of the relevant ${\cG}/\cH$.
\par
This equation is a part of the 
 Maurer-Cartan equation 
\begin{equation}
\label{eq:Maurer-Cartan}
D {\mathcal {V}}_{AB} =
\tfrac{1}{2}\mathcal{P}_{ABCD}\overline{\mathcal{V}}^{\, CD}\, ,  
\end{equation}
where $D$ is the $\cH$-covariant derivative and
$\mathcal{P}_{ABCD}$ the vielbein 1-form on the scalar manifold. Using the
definition of the graviphoton field strength (\ref{eq:graviphotondef}) we also find that
\begin{equation}
D T_{AB} =
\tfrac{1}{2}\mathcal{P}_{ABCD}\wedge \overline{T}^{\, CD}\, ,  
\end{equation}
and its complex conjugate.

It is useful in the context of black holes to define the central charges, as integrals over the dressed, scalar dependent graviphoton, $Z_{AB}$ and $\bar Z^{AB}$
and symplectic doublet charges $Q$ which are integrals over field strength's $F$ and $G$ which are scalar independent. These are related as follows
\begin{equation}
 \frac{1}{2} Z_{AB} \bar Z^{AB} =-\frac{1}{2} Q^t \cM (\cN)
 Q\,, \label{sumrule}
\end{equation}
where $\mathbb{C}$ is the symplectic metric while $\cM(\cN)$ and
$Q$ are:
\begin{eqnarray}
\cM (\cN) &=& \left( \begin{array} {cc} \bfone & - {\rm Re} \cN \cr 0 &\bfone\cr \end{array} \right )
\cdot   \left( \begin{array} {cc} {\rm Im} \cN & 0 \cr 0 &{\rm Im} \cN^{-1}\cr \end{array} \right ) \cdot
\left( \begin{array} {cc} \bfone
& 0 \cr - {\rm Re} \cN & \bfone \cr \end{array} \right ) 
=  \mathbb{C}\,{\bf V}\,{\bf V}^\dagger\,\mathbb{C}\,,\nonumber\\&&
\label{m+}
\end{eqnarray}
\be
Q=
\left(\begin{array}{c} p^\Lambda \cr q_\Lambda \cr \end{array}\right).
\ee
More useful relations follow
 \begin{eqnarray}
{\bf f}\,{\bf f}^\dagger &= &-{\rm i} \left( \cN - \bar \cN \right)^{-1}\,,\nonumber \\
{\bf h}\,{\bf h}^\dagger &= &-{\rm i} \left(\bar \cN^{-1} -
\cN^{-1} \right)^{-1}\equiv
-{\rm i} \cN \left( \cN - \bar \cN \right) ^{-1}\bar \cN \,,\nonumber\\
{\bf h}\,{\bf f}^\dagger &= & \cN {\bf f}\,{\bf f}^\dagger\,,\nonumber  \\
{\bf f}\,{\bf h}^\dagger & = & {\bf f}\,{\bf f}^\dagger \bar
\cN\,.
\end{eqnarray}

\section{Integrability of the deformed twisted self-duality in $\N \geq 5$ models \label{solve_twisted_SD} }

%\subsection{Identities}
In this appendix we use matrix-like notation and omit the Lorentz, $\cG$, $\cH$ indices. 
\begin{align}
&\f^{-1}\df^{-1}=\df^{-1}\f^{-1}=i(\cN-\bar{\cN}),\\
&X\bar{X}=\lambda^2\f^{-1}\D\f\df^{-1}\bar{\D}\df=\lambda^2\f^{-1}\D \M\bar{\D}\bar{\M}\f,\\
&(X\bar{X})^n=\lambda^{2n}\f^{-1}(\D\M\bar{\D}\bar{\M})^n\f,\\
&X(\cN-\bar{\cN})=\lambda\f^{-1}\D \f(\cN-\bar{\cN})=-i\lambda\f^{-1}\D \f(\f^{-1}\df^{-1})=-i\lambda\f^{-1}\D\df^{-1}.
\end{align}

%\subsection{Action}
We consider the following action,
\begin{align}
\mathcal{L}=\alpha F^+(1-X\bar{X})^{-1}X(\cN-\bar{\cN})F^-+\beta F^+(1-X\bar{X})^{-1}({\cN}-X\bar{X}\bar{\cN})F^++{\rm h.c.},
\end{align}
where $\alpha$ and $\beta$ are complex constants. We rewrite this action by using the identities. The first term can be rewritten as
\begin{align}
&\alpha F^+(1-X\bar{X})^{-1}X(\cN-\bar{\cN})F^-\nonumber\\
&=\alpha F^+\sum_{n=0}(X\bar{X})^nX(\cN-\bar{\cN})F^-\nonumber\\
&=-i\alpha F^+\sum_{n=0}\lambda^{2n+1}\f^{-1}(\D\M\bar{\D}\bar{\M})^n\f\times \f^{-1}\D\df^{-1}F^-\nonumber\\
&=-i\alpha F^+\sum_{n=0}\lambda^{2n+1}\f^{-1}(\D\M\bar{\D}\bar{\M})^n\D\df^{-1}F^-.
\end{align}
The hermitian conjugate of this action is
\begin{align}
&i\bar{\alpha} F^-\sum_{n=0}\lambda^{2n+1}\df^{-1}(\bar{\D}\bar{\M}\D\M)^n\bar{\D}\f^{-1}F^+\nonumber\\
&=i\bar{\alpha}F^+\sum_{n=0}\lambda^{2n+1}\f^{-1}(\D\M\bar{\D}\bar{\M})^n\D\df^{-1}F^-+{\rm tot.div},
\end{align}
where we have used the partial integral, and note that $\bar\D$ becomes $\D$ by raising and lowering Lorentz indices, which do not change the sign.

The second term becomes
\begin{align}
&\beta F^+(1-X\bar{X})^{-1}({\cN}-X\bar{X}\bar{\cN})F^+\nonumber\\
&=\beta F^+(1-X\bar{X})^{-1}\{({\cN}-\bar{\cN})-(1-X\bar{X})\bar{\cN})\}F^+\nonumber\\
&=\beta F^+\cN F^++\beta F^+\sum_{n=1} (X\bar{X})^n(\cN-\bar{\cN})F^+\nonumber\\
&=\beta F^+\cN F^++\beta F^+\sum_{n=1}\lambda^{2n} \f^{-1}(\D\M\bar{\D}\bar{\M})^n\f(\cN-\bar{\cN})F^+\nonumber\\
&=\beta F^+\cN F^++\beta F^+\sum_{n=1}\lambda^{2n} \f^{-1}(\D\M\bar{\D}\bar{\M})^{n-1}\D\M\bar{\D}\bar{\M}\f(\cN-\bar{\cN})F^+\nonumber\\
&=\beta F^+\cN F^++\beta F^+\sum_{n=1}\lambda^{2n} \f^{-1}(\D\M\bar{\D}\bar{\M})^{n-1}\D\M\bar{\D}\df(\cN-\bar{\cN})F^+\nonumber\\
&=\beta F^+\cN F^+-i\beta F^+\f^{-1}\sum_{n=1}\lambda^{2n} (\D\M\bar{\D}\bar{\M})^{n-1}\D\M\bar{\D}\f^{-1}F^+.
\end{align}
Therefore the Lagrangian becomes
\begin{align}
\mathcal{L}=&\beta F^+\cN F^+-i\beta F^+\f^{-1}\sum_{n=1} \lambda^{2n}(\D\M\bar{\D}\bar{\M})^{n-1}\D\M\bar{\D}\f^{-1}F^+
\nonumber\\
&+2({\rm Im}\alpha) F^+\sum_{n=0}\lambda^{2n+1}\f^{-1}(\D\M\bar{\D}\bar{\M})^n\D\df^{-1}F^-\nonumber\\
&+\bar{\beta}F^-\bar{\cN} F^-+i\bar{\beta} F^-\df^{-1}\sum_{n=1}\lambda^{2n} (\bar{\D}\bar{\M}\D\M)^{n-1}\bar{\D}\bar{\M}\D\df^{-1}F^-+{\rm tot.div}.
\end{align}
Let us recall the form of dual tensor $G$, and rewrite it with identities.
\begin{align}
G^+&=(1-X\bar{X})^{-1}[X(\cN-\bar{\cN})F^-+(\cN-X\bar{X}\bar{\cN})F^+]\nonumber\\
&=-i\sum_{n=0}\lambda^{2n+1}\f^{-1}(\D\M\bar{\D}\bar{\M})^n\D\df^{-1}F^-+\cN F^+-i \f^{-1}\sum_{n=1}\lambda^{2n} (\D\M\bar{\D}\bar{\M})^{n-1}\D\M\bar{\D}\f^{-1}F^+.
\end{align}
 On the other hand, 
 \begin{align}
 G^+=&\frac{i}{2}\frac{\partial S}{\partial F^+}\\
 =&i({\rm Im}\alpha) \sum_{n=0}\lambda^{2n+1}\f^{-1}(\D\M\bar{\D}\bar{\M})^n\D\df^{-1}F^- +i\beta \cN F^+
 \cr
 &+\beta \f^{-1}\sum_{n=1}\lambda^{2n} (\D\M\bar{\D}\bar{\M})^{n-1}\D\M\bar{\D}\f^{-1}F^+.
 \end{align}
 Note that we have performed partial integrals, lowering and raising operations in the above.
 Thus, by choosing $\alpha=-i$ and $\beta =-i$, we can reproduce the deformed dual tensor $G$ from the action.

%\subsection{Checking }

We would like to check the integrability condition \rf{int1} more carefully. 
As a simple example consider the first (i.e. ${\cal O}(\lambda^2)$) correction discussed in sec.~\ref{exampleLaSq}, but dressed with scalars:
\begin{align}
\mathcal{L}= - i F^+X\bar{X}({\cN}-\bar{\cN})F^++{\rm h.c.}= - i \lambda^2 F^+ \f^{-1} \D\M\bar{\D}\f^{-1}F^+ +{\rm h.c.}.
\label{action}\end{align}
Here
 \begin{align}
f^{\Sigma}{}_{AB}(\bar{f}^{-1})_{CD\Sigma}=M_{ABCD}.
 \end{align}
From the identity,  the conjugate to eq.\rf{iden} we get
\begin{align}
(\bar{f}^{-1})_{CD\Sigma}=i(\cN-\bar{\cN})_{\Sigma\Lambda}f^{\Lambda}{}_{CD},
\end{align}
and
\begin{align}
M_{ABCD}=if{}^{\Sigma}{}_{AB}(\cN-\bar{\cN})_{\Sigma\Lambda}f^{\Lambda}{}_{CD}.
\end{align}
Then, the part of the action can be written as
\begin{align}
iF^+(f^{-1})^{AB}\Delta f^{\Sigma}{}_{AB}(\cN-\bar{\cN})_{\Sigma\Lambda'}f^{\Lambda'}{}_{CD}\bar{\D}(f^{-1})^{CD} F^{+}
\label{indices}
\end{align}
and
 \begin{align}
F^+(f^{-1})^{AB}
\, \Delta \, M_{AB\, CD} \, \bar{\D}\, (f^{-1})^{CD} F^{+}.
\label{action}
 \end{align}
Note that on an $\cH$-invariant the covariant derivative is a simple one:
 \begin{align}
D S= d S.
 \label{sc}\end{align}
For $S= K_{AB} \bar K^{AB}$ we find that
 \begin{align}
D K_{AB}= d K_{AB} + B_{[A}{}^C K_{CB]}
 \end{align}
and
 \begin{align}
\bar D \bar K^{AB}= d \bar K_{AB} + \bar B^{[A}{}_C \bar K^{CB]}.
 \end{align}
To agree with $D S= d S$ we need the $\cH$-connection to be antihermitian
 \begin{align}
B= - B^\dagger.
 \end{align}
Now we present (\ref{action}) as follows
 \begin{align}
\tilde F^{AB}
\, \overrightarrow {\Delta} \, \tilde M_{AB}
\label{simpleaction}
 \end{align}
 where we have defined
  \begin{align}
\tilde M_{AB}  \equiv  M_{AB\, CD} \, \bar{\D}\, (f^{-1})^{CD} F^{+}  \qquad  \tilde F^{AB}   \equiv iF^+(f^{-1})^{AB}
\label{def}
 \end{align} 
since we are only interested in $\cH$-covariant properties. We perform partial integration in (\ref{simpleaction}) and use the fact that $d$ becomes $-d$ and our $\Delta$ has 2 factors $d+B$, each becomes $-d + B^T$ to act to the left. We use the antihermitian property of $B$ and replace it by $-d - \bar B$. Since $\Delta$ has 2 of these factors we find that
 \begin{align}
\tilde F^{AB}
\, \overrightarrow {\Delta} \, \tilde M_{AB} = \tilde F^{AB}
\, \overleftarrow {\bar \Delta} \, \tilde M_{AB}.
\label{action2}
 \end{align}
The action acquires a form
 \begin{align}
iF^+(f^{-1})^{AB} \overleftarrow {\bar \Delta}
 \, M_{AB\, CD} \, \overrightarrow{\bar \D}\, (f^{-1})^{CD} F^{+}.
\label{actionF}
 \end{align}
This is a confirmation of a consistency condition at this level. In the linear approximation it  gives a local amplitude which has at least 6 points 
\be
\langle h^{+ +} \, h^{+ +}\, h^{- -}\, h^{- -}\, v^{+ }\, v^{+ } \rangle + {\rm h.c}.
\ee
and more. But is also seems to  hint towards some kind of $U(1)$ anomaly
\be
\langle h^{+ +} \, h^{+ +}\, h^{- -}\, h^{- -}\, v^{+ }\, v^{+ } \rangle - {\rm h.c.}.
\ee
We know from  \cite{Marcus:1985yy} that the $U(1)$ subgroup in $H=U(5)$ in $\cN=5$ and $H=U(6)$ in $\cN=6$ are anomaly free, in $H=SU(8)$ in $\cN=8$ there is no $U(1)$ subgroup. Moreover, it was established more recently that there is no one-loop anomaly in $\cN\geq 5$ supergravities.

\section{ Examples of symplectic sections $(f,h)$ \label{EGsymplecticsections}}

%\subsection{ $\cN=5$ supergravity }
The action and supersymmetry rules of $\cN=5$ supergravity were given in \cite{deWit:1981yv}. The symplectic sections were presented  in
\cite{Ferrara:2008ap}, and we refer to notations and details in \cite{Ferrara:2008ap}. The theory has  5 complex scalars  $z^i$, and $\Lambda=ij$ and the symplectic section is:
 \begin{equation}
f_{\phantom{ij}AB}^{ij}=\left( e_{1}\delta _{AB}^{ij}+\frac{e_{1}}{2}%
\epsilon ^{ij ABm}z_{m}+2e_{2}\delta _{\lbrack i}^{[A}z^{B]}z_{j]}\right) \, , \qquad i=1,2,3,4,5\ ,
\label{Wed-9}
\end{equation}

\begin{equation}
h_{ij\mid AB}=\mathcal{N}_{ij\mid mn}f^{mn}{}_{AB},  \label{Wed-10}
\end{equation}

\begin{equation}
\mathcal{N}_{ij\mid kl}=-\frac{i}{1-(z_{m})^{2}}\left( \frac{1}{2}\left[
1+(z_{n})^{2}\right] \delta _{kl}^{ij}-\frac{1}{2}\epsilon
^{ijklp}z_{p}-2\delta _{\lbrack i\,[k}z_{l]}z_{j]}\right) \ ,  \label{Wed-7}
\end{equation}

\begin{eqnarray}
h_{ij\mid AB}= -i \left[ \frac{e_{1}}{2}\delta _{AB}^{ij}-\frac{e_{1}}{4}\epsilon
^{ijABk}z_{k}+e_{2}\delta _{\lbrack i}^{[A}z^{B]}z_{j]}\right] .
\label{Wed-11}
\end{eqnarray}
Here $e_{1}^2\equiv{1\over 1-|z|^2}$, $e_{2}\equiv {1-e_{1}\over |z|^2}$.

\

%\subsection{ $\cN=8$ supergravity }
The action and supersymmetry rules of $\cN=8$ supergravity were given in \cite{Cremmer:1979up} and in 
\cite{deWit:1982bul}. Here we are using the ones in \cite{deWit:1982bul},  in $SL\left( 8,\mathbb{R}\right)$-basis.   
The translation between  between the symplectic
formalism for extended supergravities reviewed in \cite
{Andrianopoli:2006ub} and the original formulation of $\mathcal N=8$
supergravity of \cite{deWit:1982bul}, (including the  more recent analysis of the gauge-fixing local $SU(8)$ in \cite{Kallosh:2008ic}), which was presented in \cite{Ceresole:2009jc}.

The coset representative for $E_{7(7)}/SU(8)$ was parametrized in \cite{deWit:1982bul} as follows
\bea\label{} \mc V&=& \left(
\begin{array}{cc}
u_{ij}^{IJ}&v_{ijKL}\\
v^{klIJ}&u^{kl}_{KL}\, 
\end{array}
\right) .\eea The sub-matrices $u$ and $v$ carry indices of  both
$E_{7(7)}$ and $SU(8)$ ($I=1,\ldots,8$, $I=1,\ldots,8$)
but one can choose a {\it suitable $SU(8)$ gauge for the fields, and
then retain only manifest invariance with respect to the rigid
diagonal subgroup of $E_{7(7)}\times SU(8)$, without distinction
among the two types of indices.} Comparing the notation of \cite{deWit:1982bul} (in
particular the appendix B) with the symplectic formalism of \cite
{Gaillard:1981rj,Andrianopoli:2006ub}, we can identify
\bea\label{} \left\{
\begin{array}{c}
\phi_{0}\equiv u\nn\\
\phi_{1}\equiv v
\end{array}
\hskip 2 cm
\begin{array}{l}
u_{ij}^{\ph{ij}kl}=(P^{-1/2})_{ij}^{\ph{ij}kl}\ ,\\
v^{ijkl}=-(\bar P^{-1/2})^{ij}_{\ph{ij}mn}\bar y^{mnkl}
\end{array}\right.
\eea so that \ba \left\{
\begin{array}{c}
 f=\frac1{\sqrt2}(\phi_{0}+\phi_{1})=\frac1{\sqrt2}(u+v)\\
i h=\frac1{\sqrt2}(\phi_{0}-\phi_{1})=\frac1{\sqrt2}(u-v)
\end{array}
\right.\ . \ea Since sections are sub-matrices of the symplectic
representation, relatively to electric and magnetic subgroups, their
explicit indices components are given by \ba\label{sections}
f_{ij}^{\ph{ij}kl}=\frac1{\sqrt2}\left(
(P^{-1/2})^{\ph{ij}kl}_{ij}-(\bar P^{-1/2})^{ij}_{\ph{ij}mn}\bar
y^{mnkl}
\right)\ ,\nn\\
h_{ij,kl}=\frac{-i}{\sqrt2}\left( (P^{-1/2})^{\ph{ij}kl}_{ij}+(\bar
P^{-1/2})^{ij}_{\ph{ij}mn}\bar y^{mnkl} \right)\ , \ea
where, in matrix notation,
\ba\label{} P=1-YY^{\dag}\
,\qquad Y=B\frac{\tanh \sqrt{B^{\dag}B}}{\sqrt{B^{\dag}B}}\ , \qquad
B_{ij,kl}=-\frac{1}{2\sqrt2}\phi_{ijkl}\ , \ea the last definition
coming from the choice of the symmetric gauge for the coset
representative in eq. (B.1) of \cite {deWit:1982bul}. If one defines
\ba\label{} \tilde P=1-Y^{\dag}Y\ , \ea and uses
the identity
\ba\label{} (\tilde
P^{-1/2})Y^{\dag}=Y^{\dag}(P^{-1/2})\, , \ea
the following simple expressions for $\mathbf{f}$ and $\mathbf{h}$ are finally
achieved:
\ba\label{sezione f}
 f&=&\frac1{\sqrt2}\left[ P^{-1/2}-(\tilde P^{-1/2})Y^{\dag} \right]=\frac1{\sqrt2}[1-Y^{\dag}]\frac1{\sqrt{1-YY^{\dag}}}\ ,\\
\label{sezione h}
 h&=&-\frac{i}{\sqrt2}\left[P^{-1/2}+(\tilde
P^{-1/2})Y^{\dag}\right]=-\frac{i}{\sqrt2}[1+Y^{\dag}]\frac1{\sqrt{1-YY^{\dag}}}\ . \ea
The above notations  are such that \ba\label{}
P^{1/2}&=&\sqrt{1-YY^{\dag}}\qquad    P_{ij}^{\ph{ij}kl}=\delta_{ij}^{kl}-y_{ijmn}\bar y^{mnkl}\nn\\
\tilde P^{1/2}&=&\sqrt{1-Y^{\dag}Y}\qquad   \bar
P^{kl}_{\ph{kl}ij}=\delta^{kl}_{ij}-\bar y^{klmn}y_{mnij}. \ea
It is easily checked that the  symplectic sections satisfy the usual relations \ba\label{}
i(\mb{f^{\dag}h-h^{\dag}f})&=&1\ ,\nn\\
\mb{h}^{T}\mb f-\mb f^{T}\mb h&=&0\ . \ea These are obtained writing
the symplectic sections as in (\ref{sezione f}) and (\ref{sezione
h}), and using the identity \ba Y\tilde{P}^{-1}=P^{-1}Y\ . \ea
The kinetic matrix is given in terms of the symplectic sections by
\cite{Andrianopoli:2006ub} \ba\label{} \mc N&=&\mb h\mb f^{-1}\ .\ea
Therefore, eqs. (\ref{sezione f}) and (\ref{sezione h}) yield
\ba\label{}
\mc N&=&-i\,[1+Y^{\dag}]\frac1{\sqrt{1-YY^{\dag}}} \sqrt{1-YY^{\dag}}\frac1{1-Y^{\dag}}=\nn\\
&=&-i\,\frac{1+Y^{\dag}}{1-Y^{\dag}}\nn
\ea
or, component-by-component,
\ba
\mc N_{ij|kl}&=&-i(\delta_{mn}^{kl}+\bar y^{mnkl})(\delta_{ij}^{mn}-\bar y^{ijmn})^{-1}\ .\
\ea

\bibliographystyle{JHEP}
\bibliography{refs}

\end{document}